\documentclass[12pt]{article}
\usepackage{graphicx}
\usepackage{enumerate}
\usepackage{amsmath}
\usepackage{amsfonts}
\usepackage{amssymb}
\usepackage{amsthm}
\usepackage[dvipsnames]{xcolor}
\usepackage[
  bookmarks=false,
  pdfpagelabels=false,
  hyperfootnotes=false,
  hyperindex=false,
  pageanchor=false,
  colorlinks,
]{hyperref}

\newtheoremstyle{dotless}{}{}{}{}{\bfseries}{}{ }{}
\theoremstyle{dotless}

\newcommand{\bt}{\begin{theorem}}

\newcommand{\et}{\end{theorem}}
\newcommand {\be}{\begin{equation}}
\newcommand {\ee}{\end{equation}}

\newcommand{\noi}{\noindent}

\def \qed {\hfill \vrule height6pt width6pt depth0pt}
\def \noi{\noindent}

\def\seq{\subseteq}

\newcommand{\bea}{\begin{eqnarray}}
\newcommand{\eea}{\end{eqnarray}}

\def \spec#1 {\mathop{#1}}

\def \b #1 {\bf #1}
\newcommand {\bash}{\setminus}

\newcommand{\fal}{\forall}

\def \qed {\hfill \vrule height6pt width 6pt depth 0pt}
\newcommand{\ben}{\begin{eqnarray*}}
\newcommand{\een}{\end{eqnarray*}}

\newtheorem{lemma}{Lemma}
\newtheorem{theorem}[lemma]{Theorem}
\newtheorem{corollary}[lemma]{Corollary}

\def \bt{\begin{theorem}}
\def \et{\end{theorem}}
\def \bel{\begin{lemma}}
\def \enl{\end{lemma}}
\def \bc{\begin{corollary}}
\def \ec{\end{corollary}}
\def \be{\begin{equation}}
\def \ee{\end{equation}}
\def \bd{\begin{df}}
\def \ed{\end{df}}

\def \ep{\epsilon}

\def \text{\mbox}

\def \noi{\noindent}

\def \geq{\ge}
\def \leq{\le}

\def\bta{\beta}

\def\Gam{\Gamma}

\def\sbt{\subset}

\def\bta{\beta}

\def\Gam{\Gamma}

\newcommand{\si}{\sigma}

\newcommand{\bdsc}{\begin{description}}
\newcommand{\edsc}{\end{description}}

\newcommand{\dom}{{\succ}}

\def\bta{\begin{tabular}}

\def\ba{\begin{array}}
\def\ea{\end{array}}

\def \b{x_2}

\def \tv {\tilde V}
\newtheorem{df}{\sc Definition}
\def \bd{\begin{df}}
\def \ed{\end{df}}
\newcommand{\R}{\mathbb{R}}

\begin{document}

\thispagestyle{empty}

\title{A look back at the core of games in characteristic function form: some new axiomatization results{\footnote{Declarations: I have no declarations to make with regard to funding or conflicts of interest/competing interests etc. In particular, this research did not receive any specific grant from funding agencies in the public, commercial or the non-profit sectors.}}}

\author{Anindya Bhattacharya\\ Department of Economics and Related Studies, University of York,\\ York, YO10 5DD, UK\\ \url {anindya.bhattacharya@york.ac.uk}}

\date{\today}

\maketitle

\begin{abstract}

The main contribution of this paper is to provide three new results axiomatizing the core of games in characteristic function form (not necessarily with transferable utility) obeying an innocuous condition (that the set of individually rational pay-off vectors is bounded). One novelty of this exercise is that our domain is the {\em entire} class of such games: i.e., restrictions like ``non-levelness" or ``balancedness" are not required.

\end{abstract}

\vskip2em

\noi JEL Classification No.: C71.

\newpage

\section{Introduction}

For analyzing coalitional behaviour, (cooperative) games in characteristic function form with not necessarily transferable utility (i.e., with ``non-transferable utility" or NTU) is a canonical framework. Naturally, the core, as a natural set-valued prediction or ``solution" for such games received quite an amount of justifiable attention (see, e.g., among others, Kannai, 1992 or part A of Mertens and Sorin, 1994). Also, quite naturally, like several other solutions within such framework, the core has been analyzed from axiomatic standpoint (Peleg, 1985; Keiding, 1986; Nagahisa and Yamato, 1992; Hwang and Sudholter, 2001; Hwang, 2006 and most recently, Pablo Arribillaga, 2016).    

However, apart from Keiding (1986), most other authors have axiomatized the core as a solution concept {\em{only on the class of NTU games which satisfy the ``non-levelness'' condition}} (we have recalled this condition in the following section). This is possibly owing to the fact that NTU games that do not satisfy this condition may not obey many well-known and useful consistency properties like Davis-Maschler consistency, Moulin consistency etc. However, the ``non-levelness'' condition may not be very appealing {\em especially in the context of games without transferable utility} because, with this condition, if a pay-off vector is weakly dominated, then it is strongly dominated as well. Thus, by this condition, some transferability of pay-off is
implicitly smuggled in even in the set-up of non-transferable utility! Moreover, at least an important class of games--the hedonic games--which is being studied extensively over the last few years (especially in the context of matching models) does not necessarily satisfy the non-levelness condition.

So, in this context, we provide three axiomatic characterizations of the core as a solution on the {\em general} class of games in characteristic function form obeying only an innocuous condition: that the set of individually rational pay-off vectors is bounded.

We would like to mention another possibly interesting feature of our characterization results. While, especially Peleg (1985, 1992) characterized the core by consistency and converse-consistency-like axioms, the novelty of Keiding's approach was to provide an axiomatization result without invoking consistency at all. Bhattacharya (2004) and then Llerena and Rafels (2007) took a mixed approach (in a transferable utility set-up): they used both consistency-like axioms as well as axioms like ``antimonotonicity" used by Keiding (1986). This paper adopts a similar mixed approach. 

The following section gives the preliminary definitions and notation. The main axioms and some discussions of these are given in Section 3. Section 4 discusses the main axiomatization results. Next we explore the implications of a consistency axiom--a variant of the one we have used for the main two characterization results--in Section 5. We provide a concluding remark in the final section.

\section{Preliminary definitions and notation}

Let $U$ be a set of potential players that may be finite or infinite. For a set $A$ we shall denote the cardinality of $A$ by $|A|$ and the proper subset relation is denoted by $\sbt.$ For any finite subset $S$ of $U,$ by ${\R}^S$ we denote the set of all functions from $S$ to ${\R},$ the set of real numbers. We would think of elements of ${\R}^S$ as $|S|-$dimensional vectors whose coordinates are indexed by the members of $S.$\\

\bd
A {\bf{(cooperative) game in characteristic function form (without side payments)}} or an NTU game is a pair $(N,V)$ where $N$ is a finite subset of $U$ and $V$ is the (characteristic) set function which assigns to every coalition $S \seq N$ a set $V(S)$ such that:\\
(1) $V(\emptyset)~=~\emptyset;$\\
(2) For all non-empty $S \seq N,$ $V(S)$ is a non-empty proper subset of ${\R}^S;$ \\
(3) For all non-empty $S \seq N,$ $V(S)$ is closed in ${\R}^S;$\\
(4) For all non-empty $S \seq N,$ $V(S)$ is comprehensive, i.e., if $x \in V(S),$ and $y \in {\R}^S$ is such that $y \leq x$ then $y \in V(S).$\\
\ed

For the rest of this paper we shall often refer an NTU game simply as a game with no possibility of confusion.\\

For any $N \sbt U$ and any vector $x \in {\R}^N$ we shall denote the $i$-th component of it by $x_i$ and the $S$ coordinates of it (where $S \seq N$) by $x_S.$ A non-empty subset of $N$ is called a coalition. For any two vectors $a,b \in {\R}^S$ for some $S \seq N,$ if $a_i>b_i$ for all $i \in S$ then we shall denote that as $a \gg b.$ Given a set $A \seq {\R}^S;$ (with $S \seq N$) the boundary of $A$ is denoted by $\bar{A},$ the interior of $A$ by $\dot{A}.$\\

\bd
Recall that a game $(N,V)$ is said to be of {\em transferable utility} or a TU game if for every non-empty $S \seq N$ there exists a real number $v(S) \in {\R}$ such that $V(S)=\{x \in {\R}^S|\sum_{i \in S}x_i \leq v(S)\}.$\\ 
\ed

{\indent}We assume henceforth that every game $(N,V)$ we consider satisfies the following regularity condition (see, e.g., Scarf, 1967).\\

\noindent {\bf{C1}} (Boundedness of Individually Rational Pay-off Vectors): For any $j \in N,$ let $b_j:=$$max$$\{x| x \in V(\{j\})\}.$ For all $S \seq N,$ the set $\{x \in V(S)|x_j \geq b_j$ for all $j \in S\}$ is bounded.\\

\noi Another regularity condition, as we mentioned in the previous section, imposed frequently while analyzing the class of NTU games is the following (see, e.g., Aumann, 1985; Peleg, 1985, 1992 etc).\\

\noindent {\bf{C2}} (Non-levelness): If $x\in \bar{V}(S)$ then $y\in {\R}^S$ and  [$y\leq x~;y \neq x$] imply that $y~\in \dot{V}(S).$\\

\noi But {\em{we shall not impose this condition}} (apart from in part ($ii$) of Theorem 2--for illustrating a minor contrasting result). Recall that every TU game, in particular, satisfies C2 and we repeat: if a general NTU game satisfies C2 then some transferability of pay-off is implicitly smuggled in even in the set-up of non-transferable utility.

\noi By $\Gam$ we denote the class of games which satisfy C1.\\

\noi A vector $x \in {\R}^T,~(T \seq N,~T \neq \emptyset)$ is said to be blocked or dominated  by a vector $y$ if there is a coalition $S \seq T$ such that $y_i>x_i$ for all $i \in S$ and $y_S \in V(S)$. We indicate this domination relation as $y \succ_S x$, i.e., $y$ dominates $x$ via coalition $S$. If vector $y$ dominates a vector $x$ via some coalition $S$ then we shall denote that as $y \succ x.$

\bd
The {\bf{core}} of the game $(N, V),$ denoted by $C(N,V)= \{x \in V(N)|$
there is no $y$ such that $y \succ x\}.$\\
\ed
The set of Pareto-efficient pay-off vectors of $(N,V),$ denoted by $X(N,V)= \{x \in V(N)|$ there is no $y$ such that $y$ dominates $x$ via $N\}.$ The set of individually rational pay-off vectors of $(N,V),$ denoted by $I(N,V)= \{x \in V(N)|$ $x_i \geq b_i$ for all $i \in N\}$ (where the notation $b_i$ has been defined in the statement of C1 above).
  
\bd Given some $\Gam_0 \seq \Gam,$ a {\bf {solution}} on $\Gam_0$ is a mapping $\si$ which
associates with each game $(N,V) \in \Gam_0$ a subset $\si (N,V)$ of
$V(N).$ \ed 

\noi We repeat that almost all of our exercises are on the entire $\Gam$ which is a central feature of this paper.\\

\bd For a game $(N,V),$ a {\bf{subgame}} of $(N,V)$ on $T \seq N,$ denoted by $(T,V_T),$ is defined as\\
\begin{center}
for all $S \seq T,$ $V_T(S)=V(S).$
\end{center}
\ed

\section{The main axioms}

\noi Several of the axioms used in this paper are quite well-known in literature and hardly need much discussion. Below we discuss in some detail the axioms we consider relatively less familiar.\\

\noi We state the axioms by invoking an arbitrary solution $\si$ on the general domain of our study: the entire $\Gam;$ but restating these on some subsets of $\Gam,$ if required, is unproblematic. Take $(N,V) \in \Gam.$\\

{\noindent}1. {\bf{Pareto Optimality}} (PO): $\si (N,v) \seq X(N,V).$\\

{\noindent}2. {\bf{Non-emptiness for Single Player Games}} (NESPG): For every game $(N,V)$ with $|N|=1,$  $\si (N,V) \neq \emptyset.$\\

{\noi}3. {\bf{Irrelevance of $\si$-empty Coalitions}} (IREC):\\

\noi Suppose that for every non-singleton and non-empty $S \sbt N,$ $\si(S,V_S)=\emptyset.$ In that case, if there exists $x \in I(N,V),$ then $\si(N,V) \neq \emptyset.$\\

\noi The idea behind this axiom is that if a coalition is to affect the solution set for the grand coalition, then it must have a non-empty solution set for its own sub-situation (i.e., the respective subgame). Operationally this is a quite weak ``non-emptiness" axiom. It is straightforward that if a game satisfies IREC then that satisfies NESPG as well.\\

\noi Next we provide some consistency-like axioms and toward that goal we define a reduced game.

\bd Let $x \in X(N,V).$ The {\bf{strong secession reduced game}} (SS reduced game hereafter) on $S \sbt N,$ $(S \neq \emptyset)$
with respect to $x,$ $(S,V^x_S),$ is given by:

\[ V^x_S(S)=
                            \left\{ \begin{array}{ll}
                                    V(S)  & \mbox{if $x_S \in V(S)$,}\\
                                    \{y \in {\R}^S|~y \leq x_S\} & \mbox{if $x_S \not\in V(S),$}\\

                                   \end{array}
                             \right.  \]

$$ V^x_S(T)=V(T)~{\mbox{for}}~T \sbt S.$$
\ed

This reduced game reflects the following situation. Suppose, a pay-off vector $x$ is agreed upon by $N.$ Then, the players in $N \bash S$ leave, no cooperation with them is possible any more. But the grand coalition $S$ in the ``reduced'' situation can still renegotiate on the pay-off distribution. If it finds that it cannot possibly improve upon this agreed pay-off $x$ then this distribution is maintained in the reduced situation as well. Otherwise, the members of $S$ oppose the pay-off distribution according to $x$ and completely secede from the original game making a coalition for themselves. Moreover, since no cooperation with the players in $N \bash S$ is possible, the worth of each $T \sbt S$ in the reduced game remains what it was in the original
game. This is similar in spirit to the reduced game introduced by Nagahisa and Yamato (1992). Bhattacharya (2004) used such a reduced game in transferable utility set-up. Llerena and Rafels (2007) also used this idea (but they called it ``projection" reduced game).\\

{\noindent}4. {\bf{Strong Secession Consistency}} (SSC):\\
If $x \in \si(N,V)$ then for any coalition $S,$ $x_S \in \si(S,V^x_S).$\\

\noi The corresponding ``converse" property is:\\

{\noindent}5. {\bf{Converse Strong Secession Consistency}} (CSSC):\\
Suppose $x \in X(N,V)$ and for every coalition $S,~S \neq N,$ $x_S \in \si(S,V^x_S).$ Then $x \in \si(N,V).$\\

\noi The next axiom is akin to continuity (but weaker than continuity) which is a desirable feature for a solution. For stating this axiom, to represent distance between two subsets of a finite-dimensional Euclidean space we have used below the Hausdorff distance as that is possibly the most widely used in such contexts.\\

{\noindent}6. {\bf{Weak continuity}} (WC):\\
Let $\{(N,V^k)\}$ be a sequence of games such that $\fal k,$ $V^k(S)=V(S)$ for $S \sbt N$ and $V^k(N)$ converges to $V(N)$ (in the Hausdorff distance).  Let $\{x^k\}$ be a sequence such that $x^k \in \si(N,V^k)$ for all $k$ and $x^k$ converges to $x.$ Then $x \in \si(N,V).$\\

{\noi}7. {\bf{Antimonotonicity}} (AM):\\
Let $(N,V') \in \Gam$ be such that $V'(S) \seq V(S)$ for all $S \sbt N$ and $V'(N)$$=$$V(N).$ Then $\si(N,V) \seq \si(N,V').$\\

{\indent}The intuition is that if the coalitions get impoverished
then the pay-off vectors in the solution of the original game
remain in the solution of the new game and additionally some more
pay-off vectors feasible for the grand coalition may qualify as
solution vectors. Keiding (1986) introduced this axiom in the literature.\\

It is easy to see that the core satisfies each of the above axioms on $\Gam.$

\section{The main characterization results}

{\noindent}{\bf{Theorem 1}} {\em{There is a unique solution on $\Gam$ that satisfies PO, NESPG, SSC and CSSC and it is the core. Further, these four axioms are independent on $\Gam:$ i.e., for each of these axioms there exists a solution which, on $\Gam,$ violates this axiom but satisfies the other three.}}\\

For proving the characterization part of this Theorem we use the following two lemmas. This proof is similar to the proof of a somewhat similar result in Nagahisa and Yamato (1992).\\

{\noindent}{\bf{Lemma 1.1}} {\em{
If a solution $\si(.)$ satisfies PO and SSC on $\Gam$ then for any $(N,V) \in \Gam,$ $\si(N,V) \seq C(N,V).$}}\\

{\noi}{\bf{Proof.}} Take $(N,V) \in \Gam$ and suppose that $x \in
\si(N,V) \bash C(N,V).$ Then, for some $S \sbt N,$ $x_S \in
\dot{V}(S).$ Therefore, by the definition of a SS reduced game,
$V^x_S(S)=V(S).$ Moreover, by SSC, $x_S \in \si(S,V^x_S).$ But
since $x_S \in \dot{V}(S),$ then $\si(.)$ violates PO.\qed\\

{\noindent}{\bf{Lemma 1.2}} {\em{
If a solution $\si(.)$ satisfies PO, NESPG and CSSC on $\Gam$ then for any $(N,V) \in \Gam,$ $C(N,V) \seq \si(N,V).$}}\\

{\noi}{\bf{Proof.}} We shall prove this by induction on the number of players. Take
 $(N,V) \in \Gam.$ If $|N|=1,$ then by NESPG, $\si(N,V) \neq \emptyset$ and by PO,
  $C(N,V) \seq \si(N,V)$ (in fact, these two sets are equal). Assume that the result is true whenever $|N|$ is less than or equal to
   some positive integer $k-1.$ Consider $(N,V) \in \Gam$ such that $|N|=k.$ Let $x \in C(N,V).$
    Then, for every $S \sbt N,$ $x_S \in C(S,V^x_S)$ and therefore, by the induction hypothesis,
     $x_S \in \si(S,V^x_S).$ Then, by CSSC, $x \in \si(N,V).$\qed\\

This completes the proof of the characterization part.\\

\noi {\bf{Proof of the remainder of Theorem 1}}\\

\noi Next we show that each of the axioms are independent of the others.\\

\noi PO: Consider a solution $\si$ on $\Gam$ as follows:\\
if a game $(N,V)$ is a balanced{\footnote{For the definition of a balanced TU game, see, if required, Kannai (1992). And for the piece of notation $v(S)$, here and later in the paper, please refer to Definition 2 in Section 2 above.}} TU game then $\si(N,V)=\{x \in V(N)|$ for every non-empty proper subset $S$ of $N,$ $\sum_{i \in S}x_i \geq v(S)\}$ and $\si(N,V)$ is $C(N,V)$ otherwise. Then $\si$ violates PO but satisfies each of the other three axioms for this Theorem.\\

\noi NESPG: Consider a solution $\si$ on $\Gam$ as follows: for every $(N,V) \in \Gam,$ $\si(N,V)=\emptyset.$ Then $\si$ violates NESPG but satisfies each of the other three axioms for this Theorem.\\

\noi SSC: Consider a solution $\si$ on $\Gam$ as follows: for every $(N,V) \in \Gam,$ $\si(N,V)=\{x \in X(N,V)|$ for no $S \subset N$ with $|S|=|N|-1$ is it the case that there exists some $y \in V(S)$ such that $y_i>x_i$ for each $i \in S\}.$\\
Then it is easy to see that $\si$ satisfies PO and NESPG but violates SSC.\\
To show that $\si$ satisfies CSSC, let $x \in X(N,V)$ be such that for every proper subset $S$ of $N,$ $x_S \in \si(S,V^x_S)$ but $x \notin \si(N,V).$ Then there exists $S \subset N$ with $|S|=|N|-1$ for which there exists some $y \in V(S)$ such that $y_i>x_i$ for each $i \in S.$ This implies $V^x_S(S)=V(S).$ But then $x_S \notin X(S, v^x_S)$ which leads to a contradiction.\\

\noi CSSC: Consider a solution $\si$ on $\Gam$ as follows: for every $(N,V)$ for which $|N|=1,$ $\si(N,V)=C(N,V)$ and if $|N|>1$ then $\si(N,V)=\emptyset.$ Then $\si$ violates CSSC (which is easy to see considering examples of games with $|N|=2$) but satisfies the other three axioms. \qed\\

\noi The proof of the next characterization result--Theorem 2--uses some ideas from Bhattacharya (2004).\\

{\noindent}{\bf{Theorem 2}} {\em{(i) There is a unique solution on $\Gam$ that satisfies PO, IREC, SSC, WC and AM and it is the core. Further, these five axioms are independent on $\Gam:$ i.e., for each of these axioms there exists a solution which, on $\Gam,$ violates this axiom but satisfies the other four.\\
(ii) However, let $\Gam_0 \subset \Gam$ be the subclass of games such that each $(N,V)$ in $\Gam_0$ satisfies C2 (i.e., non-levelness) and for each $(N,V) \in \Gam_0,$ $|N|=2.$ Then there is a unique solution on $\Gam_0$ that satisfies PO, IREC, SSC and AM and it is the core (i.e., for characterization of the core for this sub-class, WC is not required).}}\\

{\noi}{\bf{Proof.}} {\em Proof of Part (i)}: Suppose a solution $\si(.)$ satisfies PO, IREC, SSC, WC and AM on $\Gam.$
Take $(N,V) \in \Gam.$ By Lemma 1.1 above, $\si(N,V) \seq C(N,V).$\\
{\indent}Take $x \in C(N,V).$ Fix $\ep >0$ and construct the game $(N,V^\ep)$ as follows:\\
$$V^\ep (N)=V(N) \cup \{y \in {\R}^N|~\fal i \in N,~y_i \leq x_i + \ep/|N|\},$$
and for $S \sbt N,$
$$V^\ep (S)=V(S).$$
Construct the vector $x^\ep,$ given by $x_i^{\ep}=x_i + \ep/|N|$ for all $i \in N.$\\
{\indent}Now, further construct the game $(N, V^{\ep,x})$ for which
$V^{\ep,x}(S)=V^\ep(S)$ for every non-singleton coalition $S \seq N$ and for every $i \in N,$
$V^{\ep,x}(\{i\})=\{y \in {\R}|y \leq x_i + \ep/|N|\}$.\\
{\indent} We claim that for any proper coalition $S \sbt N$ such that $|S|>1,$ $\si(S, V_S^{\ep,x}),$ i.e., the solution for the subgame of $(N,V^{\ep,x})$ on the coalition $S,$ is empty. Suppose otherwise. Fix $S \sbt N,$ $|S| >1,$ such that $\si(S, V_S^{\ep,x}) \neq \emptyset$ and let $y \in \si(S, V_S^{\ep,x}).$ By Lemma 1.1, $y_i \geq x_i + \ep/|N|$ for every $i \in S.$ But then $y \dom_S x$ which contradicts the supposition that $x \in C(N,V).$ Hence, the claim is proved.\\
{\indent}Then, by IREC, $\si(N, V^{\ep,x}) \neq \emptyset$ and by PO and Lemma 1.1, $\si(N, V^{\ep,x})=\{x^\ep\}.$ Therefore, by AM, $x^\ep \in \si(N, V^{\ep}).$ Now, take a decreasing sequence of positive numbers $\{\ep^k\}$ such that $\ep^1=\ep$ and $\ep^k$ $\longrightarrow$ $0.$ For each $k,$ construct a game $(N,V^{\ep^k})$ such that:\\
$$V^{\ep^k}(N)=V(N) \cup \{y \in {\R}^N|y_i \leq x_i + \ep^k/|N|~\fal i \in N\},$$
and for $S \sbt N,$
$$V^{\ep^k} (S)=V(S).$$
Let $x^{\ep^k}$ be the vector given by $x_i^{\ep^k}=x_i + \ep^k /|N|$ $\fal i \in N.$
By our argument above, for each $k,$ $x^{\ep^k}$ is in $\si(N,V^{\ep^k}).$ Then for the sequence $\{(N,V^{\ep^k})\},$ $V^{\ep^k} (N)$ converges to $V(N)$ (in the Hausdorff distance) and $x^{\ep^k}$ converges to $x.$ Then, by WC, $x \in \si(N,V).$\\

\noi Next we show that each of the axioms are independent of the others.\\

\noi PO: Consider a solution $\si$ on $\Gam$ as follows: for every $(N,V) \in \Gam,$ $\si(N,V)=I(N,V),$ the set of individually rational pay-off vectors for $(N,V).$ Then it is straightforward to see that $\si$ violates PO but satisfies the other four axioms.\\

\noi IREC: Consider a solution $\si$ on $\Gam$ as follows: for every $(N,V) \in \Gam,$ $\si(N,V)=\emptyset.$ Then it is straightforward to see that $\si$ violates IREC but satisfies the other four axioms.\\

\noi SSC: Consider a solution $\si$ on $\Gam$ as follows: for every $(N,V) \in \Gam,$ $\si(N,V)=\{x \in X(N,V)|$ for no $S \subset N$ with $|S|=|N|-1$ is it the case that there exists some $y \in V(S)$ such that $y_i>x_i$ for each $i \in S\}.$\\
Then it is straightforward to see that $\si$ satisfies PO, WC and AM but violates SSC.\\
The proof that this $\si$ satisfies IREC proceeds as follows. First note that since each $(N,V) \in \Gam$ satisfies C1, for each such $(N,V)$ with a non-empty $I(N,V),$ every $y \in I(N,V)$ cannot be in the interior of $V(N)$ and thus, $X(N,V) \cap I(N,V) \neq \emptyset.$ Take, if possible, $(N,V) \in \Gam$ for which IREC does not hold. Take $y \in X(N,V) \cap I(N,V).$ If $|N|=2,$ then $y \in \si(N,V)$ and thus IREC is satisfied (in contradiction to the supposition). Let $|N|>2.$ Take $y \in X(N,V) \cap I(N,V).$ Since the game violates IREC, $y \notin \si(N,V).$ Therefore, for some $S \subset N$ with $|S|=|N|-1,$ there exists some $z \in V(S)$ such that $z_i>y_i$ for each $i \in S.$ If $z \in \si(S,V_S)$ then $(N,V)$ satisfies IREC (vacuously) and thus, we get a contradiction. Therefore, $z \notin \si(S,V_S).$ Therefore, for some $T \subset N$ with $|T|=|S|-1,$ there exists some $w \in V(T)$ such that $w_i>z_i>y_i$ for each $i \in T.$ Repeating the same argument, finally, there must exist $i \in N$ such that $b_i > y_i.$ But then $y \notin I(N,V)$ which leads to a contradiction.\\

\noi WC: Fix a set of players $\hat N=\{1,2,3\}.$ Fix a game $({\hat N}, {\hat V})$ as follows:\\
${\hat V}({\hat N})=\{x \in {\R}^{\hat N}|\sum_{i \in {\hat N}}x_i \leq 9\};$\\ 
for $S \sbt N$ such that $|S|=2;$ ${\hat V}(S)=\{x \in {\R}^{S}|\sum_{i \in {S}}x_i \leq 3\};$\\
for $S \sbt N$ such that $|S|=1;$ ${\hat V}(S)=\{x \in {\R}^{S}|\sum_{i \in {S}}x_i \leq 1\}$.\\
Let a solution $\si$ be as follows. For any $(N,V) \in \Gamma$ such that $({\hat N}, {\hat V})$ is a subgame of $(N,V),$ $\si(N,V)=\emptyset.$ For $({\hat N}, {\hat V}),$ $\si ({\hat N}, {\hat V})=(3,3,3).$ For any other $(N,V) \in \Gamma,$ $\si(N,V)$ is the core.\\
Then it is easy to see that $\si$ violates WC but satisfies the other four axioms.\\

\noi AM: Fix a set of players $\hat N=\{1,2\}.$ Consider the subset of games $\Gamma' \sbt \Gamma$ such that for each $(N,V) \in \Gamma',$ $N={\hat N}$ and\\
${V}(N)=\{x \in {\R}^{N}|\sum_{i \in {N}}x_i \leq 2\};$\\
$b_1=1;$ $0 \leq b_2 \leq 1.$\\
Let a solution $\si$ be as follows. For any $(N,V) \in \Gamma$ such that $N={\hat N},$ $\si(N,V)=x \in C(N,V)$ such that $x_2=b_2.$ For any $(N,V) \in \Gamma$ such that $\hat N$ is a proper subset of $N$ and for which $C({\hat N}, V_{\hat N}) \neq \emptyset,$ $\si(N,V)=\emptyset.$ For any other $(N,V) \in \Gamma,$ $\si(N,V)$ is the core.\\
Then it is easy to see that $\si$ violates AM but satisfies the other four axioms.\\

\noi {\em{Proof of Part (ii)}}: Take any 2-player game $(N,V) \in \Gam_0$ and $x \in C(N,V).$ Construct $(N, V^{x}) \in \Gam_0$ for which $V^{x}(N)=V(N)$ and for every $i \in N,$
$V^{x}(\{i\})=\{y \in {\R}|y \leq x_i\}$.\\
Then, by IREC, $\si(N,V^x) \neq \emptyset.$ Note that $C(N,V^x)=\{x\}.$ To see this consider some $y \neq x$ such that $y \in C(N,V^x).$ Then $y_i \geq x_i$ for each $i \in N$ and $y_i >x_i$ for at least one $i \in N.$ But then, since $(N,V^x)$ satisfies C2 (non-levelness), $x$ is in the interior of $V(N)$ which leads to a contradiction. Since, by Lemma 1.1, $\si(N,V^x) \seq C(N,V^x),$ $\si(N,V^x)=C(N,V^x)=\{x\}.$ Then, by AM $x \in \si(N,V).$ \qed\\

\section{A related consistency property and its implication}

\noi Note that following the idea of secession consistency, a similar consistency axiom can be introduced as follows. First we define a corresponding reduced game.

\bd Let $x \in X(N,V).$ The {\bf{weak secession reduced game}} on
$S \sbt N,$ $(S \neq \emptyset)$ with respect to $x,$
$(S,{\tv}^x_S),$ is given by:

$${\tv}^x_S(S)=\{y \in {\R}^S|~y \leq x\};$$

$$ {\tv}^x_S(T)=V(T)~{\mbox{for}}~T \sbt S.$$
\ed

\noi This, obviously, is similar in spirit to the strong secession reduced game. But here, the coalition $S$ {\em must} maintain the pay-off agreed upon by $N$ and so, the power of seceding is weaker. The corresponding consistency property is:\\

{\noindent}8. {\bf{Weak Secession Consistency}} (WSC):\\
If $x \in \si(N,V)$ then for any coalition $S,$ $x_S \in \si(S,{\tv}^x_S).$\\

\noi The corresponding ``converse consistency" condition is:\\

\noi $8^{\prime}.$ {\bf{Converse Weak Secession Consistency}} (CWSC):\\
Suppose $x \in X(N,V)$ and for every coalition $S,~S \neq N,$ $x_S \in \si(S,{\tv}^x_S).$ Then $x \in \si(N,V).$\\

\noi In this section we explore the possibility of axiomatizing the core using this related but distinct consistency condition instead of SSC. But first we make a few preliminary observations.\\

{\noindent}{\bf{Observation 1}} {\em{(i) The core does not satisfy CWSC on $\Gam;$\\
(ii) There exists a solution $\si$ on $\Gam$ which satisfies WSC but not SSC;\\
(iii) If a solution $\si$ on $\Gam$ satisfies SSC and AM then it satisfies WSC.}}\\

{\noi}{\bf{Proof.}} $(i)$: Consider the following game. $N=\{1,2,3\}$ and\\
${V}({N})=\{x \in {\R}^{N}|\sum_{i \in {N}}x_i \leq 6\};$\\ 
${V}(\{1,2\})=\{x \in {\R}^{\{1,2\}}|x_1+x_2 \leq 4\};$\\  
for every other $S \sbt N,$ ${V}(S)=\{x \in {\R}^{S}|\sum_{i \in {S}}x_i \leq 0\}$.\\
Consider the vector $x=(2,1,3).$  Then $x \in X(N,V)$ and for every coalition $S,~S \neq N,$ $x_S \in C(S,{\tv}^x_S).$ But $x \notin C(N,V).$\\
$(ii)$: Consider the following solution: for each $(N,V) \in \Gam,$ the solution is $X(N,V).$ Consider the game used in proving part $(i)$ of this Observation. For that game, $X(N,V)$ satisfies WSC but not SSC.\\
$(iii)$: Suppose a solution $\si$ on $\Gam$ satisfies SSC and AM. Take a game $(N,V)$ and $x \in \si (N,V).$ By SSC, for every $S \seq N,$ $x_S \in \si (S, V^x_S).$ Note that for every $S \seq N,$ and $T \seq S,$ ${\tv}^x_S(T) \seq V^x_S(T).$ Therefore, by AM $x_S \in \si (S, {\tv}^x_S).$ Thus, $\si$ satisfies WSC. \qed\\

\noi Next we introduce our final axiom.\\

{\noindent}9. {\bf{Weak Internal Stability for Proximal Coalitions}} (WISPC):\\
Let $x \in \si(N,V).$ Consider any $S \sbt N$ such that $|S|=|N|-1.$ Then for all $y \in \si(S,V_S),$
$$max_{j \in S}~ x_j \geq min_{j \in S}~y_j.$$

{\indent}This axiom is somewhat egalitarian in spirit. Suppose for a coalition  $S$ proximal to $N$ (obtained by dropping only one player) even the worst-paid player  in a pay-off vector $y$ in the {\em solution} of the subgame on $S$ gets more than that is given to {\em{any}} player of $S$ in an allocation $x$ for the grand coalition. Then, this axiom specifies that if $x$ is {\em{so bad}} for possibly such a {\em{large fraction of the players}} then $x$ should not be in the solution of the whole game. Bhattacharya (2004) introduced this axiom in context of transferable utility scenarios.\\

\noi Our final characterization result is as follows.\\

{\noindent}{\bf{Theorem 3}} {\em{(i) The core is the minimal among the solutions which satisfy PO, IREC, WSC, WC, WISPC and AM on $\Gamma.$\\
(ii) The minimality in the statement in part (i) is non-trivial: i.e., there exists a solution $\si$ satisfying all these six axioms on $\Gam$ such that for every $(N,V) \in \Gam,$ $C(N,V) \seq \si(N,V)$ and for some $(N,V) \in \Gam,$ $C(N,V)$ is a proper subset of $\si(N,V).$}}\\

\noi We will prove part $(i)$ of this theorem with the help of the following lemmata.\\

\noi {\bf{Lemma 3.1}} {\em{ If a solution $\si(.)$ satisfies PO,
IREC, WSC and WISPC on $\Gam$ then for any $(N,V) \in \Gam,$
$\si(N,V) \seq I(N,V).$}}\\

{\noi}{\bf{Proof.}} Take $x \in \si (N,V).$Note that by IREC, for any single-player game $(N,V),$ $\si(N,V) \neq \emptyset.$ Then, by PO, for every $i \in N,$ $\si(\{i\}, V_{\{i\}})=\{b_i\}.$
  Now, let $|N|>1$ and $x_i < b_i$ for some $i \in N.$ If $|N|=2,$ then $\si(.)$ clearly
  violates WISPC. If $|N|>2,$ then pick $j \in N \bash \{i\}$ and construct
   $(\{i,j\}, {\tv}^x_{\{i,j\}}),$ the weak secession reduced game on $\{i,j\}$ with respect to
    $x.$ Then by WSC, the vector $(x_i,x_j) \in \si(\{i,j\}, {\tv}^x_{\{i,j\}} ).$ But then again,
     $\si(.)$ violates WISPC. \qed\\

{\noindent}{\bf{Lemma 3.2}} {\em{If a solution $\si(.)$ satisfies PO, IREC, WSC, WC, AM and WISPC on $\Gam$ then for any $(N,V) \in \Gam,$ $C(N,V) \seq \si(N,V).$}}\\

{\noi}{\bf{Proof.}} The proof is exactly similar to the proof of part $(i)$ of Theorem 2 above, but for completeness we (essentially) reproduce the first part of the proof.\\
{\indent}Take $x \in C(N,V).$ Fix $\ep >0$ and construct the game $(N,V^\ep)$ as follows:\\
$$V^\ep (N)=V(N) \cup \{y \in {\R}^N|~\fal i \in N,~y_i \leq x_i + \ep/|N|\},$$
and for $S \sbt N,$
$$V^\ep (S)=V(S).$$
Construct the vector $x^\ep,$ given by $x_i^{\ep}=x_i + \ep/|N|$ for all $i \in N.$\\
{\indent}Now, further construct the game $(N, V^{\ep,x})$ for which
$V^{\ep,x}(S)=V^\ep(S)$ for every non-singleton coalition $S \seq N$ and for every $i \in N,$
$V^{\ep,x}(\{i\})=\{y \in {\R}|y \leq x_i + \ep/|N|\}$.\\
{\indent} We claim that for any proper coalition $S \sbt N$ such that $|S|>1,$ $\si(S, V_S^{\ep,x}),$ i.e., the solution for the subgame of $(N,V^{\ep,x})$ on the coalition $S,$ is empty. Suppose otherwise. Fix $S \sbt N,$ $|S| >1,$ such that $\si(S, V_S^{\ep,x}) \neq \emptyset$ and let $y \in \si(S, V_S^{\ep,x}).$ By Lemma 3.1, $y_i \geq x_i + \ep/|N|$ for every $i \in S.$ But then $y \dom_S x$ which contradicts the supposition that $x \in C(N,V).$ Hence, the claim is proved.\\
The remainder of the proof is exactly identical to that for the characterization result in part $(i)$ of Theorem 2 above. \qed\\

\noi {\bf{Proof of part $(ii)$ of Theorem 3.}} Consider the following solution $\si$ on $\Gam.$\\
If $(N,V)$ is a TU game then $\si (N,V)=\{x \in X(N,V)|$ for no $S \subset N$ with is it the case that $(v(S)/|S|)>x_i$ for each $i \in S\}$ (where, recall that $v(S)$ is as in Definition 2 above); and\\
$\si(N,V)=C(N,V)$ otherwise.\\
It is straightforward to see that $\si$ satisfies PO, WSC, WC, AM and WISPC.\\
To show that $\si$ satisfies IREC we proceed as follows. Naturally, it suffices to confine attention to $\Gamma',$ the sub-class of TU games. First it is straightforward that if $|N|=2$ then $\si$ satisfies IREC. Now suppose that for some game $(N,V),$ with $|N|>2,$ IREC is violated. If for some proper coalition $S \sbt N$ such that $|S|>1,$ $\si(S,V_S) \neq \emptyset,$ then IREC is (vacuously) satisfied. Therefore, for every proper coalition $S \sbt N$ such that $|S|>1,$ $\si(S,V_S)=\emptyset.$ This implies that if $|S|=2,$ then for some $i \in S,$ $b_i>(v(S)/|S|).$ Suppose this is true if $|S|=k \geq 2$: i.e., for every $S$ with $|S|=k,$ for some $i \in S,$ $b_i>(v(S)/|S|).$ But suppose that for some proper coalition $S \sbt N$ such that $|S|=k+1,$ $(v(S)/|S|) \geq b_i$ for every $i \in S.$ But then $(v(S)/|S|) \geq (v(T)/|T|)$ for every proper subcoalition $T$ of $S.$ But this implies that $\si (S,V_S) \neq \emptyset$ which leads to a contradiction. Now take $x \in X(N,V) \cap I(N,V).$ Then it must be that $x_S \geq (v(S)/|S|)$ for every proper coalition $S$ of $N.$ This is because, otherwise, for some $i \in N,$ $b_i > (v(S)/|S|)>x_i$ which leads to a contradiction to the supposition that $x \in I(N,V).$ Then $x \in \si(N,V).$ \qed\\     

\section{A concluding remark}

Note that while the core has an immediate intuitive explanation, some other solution concepts (like the kernel) are intuitively (apparently) less straightforward and the acceptability of these depend more strongly on axiomatic justification (e.g., Inarra et al., 2020). But much of such exercises have been under the restriction of ``non-levelnesss". Perhaps some axioms used in this paper may be fruitfully used to explore other such solutions axiomatically on richer classes of games (but, of course, at the moment this remark is entirely speculative).

\section*{Declaration of generative AI and AI-assisted technologies in the writing process}

\noi I have not used any generative AI and AI-assisted technologies at all.

\section*{Acknowledgements}

\noi For their comments and suggestions I am grateful to Zaifu Yang and especially, to Yuan Ju.

\section*{References}

\begin {description}

\item Arribillaga, R. P. (2016). ``Axiomatizing core extensions on NTU games". {\em International Journal of Game Theory}, {\bf 45}, 585–600.

\item Aumann, R. J. (1985). ``An Axiomatization of non-transferable Utility Value''. {\it{Econometrica}}, {\bf{53}}, 599-612.

\item Bhattacharya, A. (2004). ``On the equal division core". {\em Social Choice and Welfare}, {\bf 22}, 391–399.

\item Hwang, Y.-A. (2006). Two characterizations of the consistent egalitarian solution and of the core on NTU games. {\em{Mathematical Methods of Operations Research}}, {\bf{64}}, 557-568.

\item Hwang, Y.-A. and P. Sudholter (2001). ``Axiomatizations of the core on the universal domain and other natural domains''. {\em{International Journal of Game Theory,}} {\bf{29}}, 597-623.

\item Inarra, E., R. Serrano and K.-I. Shimomura, 2020. ``The Nucleolus, the Kernel, and the Bargaining Set: An Update". {\em{Revue economique,}} {\bf{71(2)}}, 225-266.

\item Kannai, Y. (1992). ``The Core and Balancedness" in {\em{Handbook of Game Theory}}, Vol. 1, (R. J. Aumann and S. Hart, Eds.), pp. 355-395. Amsterdam: Elsevier.

\item Keiding, H. (1986). ``An Axiomatization of the Core of a Cooperative Game''. {\em{Economics Letters}}, {\bf{20}}, 111-115.

\item Llerena, F. and C. Rafels (2007). ``Convex decomposition of games and axiomatizations of the core and the D-core". {\em{International Journal of Game Theory}}, {\bf{35}}, 603-615.

\item Mertens, J.-F. and S. Sorin (1994) (Eds). {\em Game-Theoretic Methods in General Equilibrium Analysis}. Springer.

\item Nagahisa, R. and T. Yamato (1992). ``A Simple Axiomatization of the Core of Cooperative Games with a Variable Number of Players''. {\em{mimeo}}: Working Paper No. 138, Department of Economics, Toyama University.

\item Peleg, B. (1985). ``An Axiomatization of the core of cooperative games without side payments''. {\it{Journal of Mathematical Economics}}, {\bf{14}}, 203-214.

\item Peleg, B. (1992). ``Axiomatizations of The Core'' in {\em{Handbook of Game Theory}}, Vol. 1, (R. J. Aumann and S. Hart, Eds.), pp. 397-412. Amsterdam: Elsevier.

\item Scarf, H. (1967). ``The Core of an N-person Game''. {\it{Econometrica}}, {\bf{35}}, 50-69.

\end{description}

\end{document}